%% file: arxiv.tex
\documentclass[a4paper, 10pt, twocolumn]{article}

\usepackage{amsfonts,amssymb,amsmath,amsthm}
\usepackage[utf8]{inputenc}
\usepackage{subfigure}
\usepackage{graphicx}
\usepackage[footnotesize]{caption}
\usepackage[setbb,setcolon]{kmath}
\usepackage[sort&compress,numbers]{natbib}
\usepackage{mathrsfs}
\usepackage[colorlinks=true,breaklinks]{hyperref}
\usepackage{cleveref}

\usepackage[top=1.5cm, left=1.5cm, right=1.5cm, bottom=1.5cm]{geometry}

\input{notations.tex}

\title{Continuous dictionaries meet low-rank tensor approximations}

\author{Clément Elvira$^1$, Jeremy E. Cohen$^2$, Cédric Herzet$^3$, Rémi Gribonval$^4$.\\
\footnotesize $^1$ Univ. Bretagne Sud,\ $^2$Univ Rennes, Inria, CNRS, IRISA,\ $^3$ Inria,\ $^4$ Univ Lyon, Inria, CNRS, ENS de Lyon, UCB Lyon 1, LIP UMR 5668, F-69342, Lyon, France.
}
\date{\empty} 

\renewenvironment{abstract}{\bf\small {\em\ Abstract---}}{}

\definecolor{green_commentary}{RGB}{35,139,69}

\begin{document}

\definecolor{myblue}{RGB}{18,75,126}

\hypersetup{
	pdfauthor={},
	pdftitle={},
	pdfsubject={},
	pdfkeywords={},
	pdfcreator={},
	pageanchor = false, 
	linkcolor  = orange!85!uiblack,
	citecolor  = myblue!100!uiblack,
	urlcolor   = green!85!uiblack,
}

\maketitle

\begin{abstract}
  In this short paper we bridge two seemingly unrelated sparse approximation topics: continuous sparse coding and low-rank approximations. We show that for a specific choice of continuous dictionary, linear systems with nuclear-norm regularization have the same solutions as a BLasso problem. Although this fact was already partially understood in the matrix case, we further show that for tensor data, using BLasso solvers for the low-rank approximation problem leads to a new branch of optimization methods yet vastly unexplored. In particular, the proposed Frank-Wolfe algorithm is showcased on an automatic tensor rank selection problem.
\end{abstract}

\newcommand{\signalSet}{\calX}
\newcommand{\obsSet}{\calH}
\newcommand{\setDic}{\calD}

\newcommand{\opObs}{\mathscr{A}}

\newcommand{\vobs}{\bfy}
\newcommand{\vcoeff}{\bfx}
\newcommand{\matDic}{\bfD}
\newcommand{\matOp}{\bfA}

\newcommand{\matobs}{\bfY}
\newcommand{\mat}{\bfX}

\newcommand{\opMat}{\mathscr{D}}
\newcommand{\mesure}{\mu}


\section{Sparse and low-rank reconstructions}

\paragraph{Sparse decompositions.} The ``$\nnz$-sparse representation problem'' consists in finding an accurate representation of some vector $\vobs\in\kR^\dimobs$ 
as the linear combination of $\nnz$ elements from some set $\setDic\subset \kR^\dimobs$, commonly called \emph{dictionary}.
 Finding the best
 $\nnz$-sparse representation of $\vobs\in\kR^\dimobs$ in $\setDic$ is known to be an NP-hard problem for most dictionaries, see~\cite[Section 2.3]{Foucart2013aa}.
 Hence, the design of tractable methods to find (provably) good sparse representations of certain families of signals $\vobs$ has been a very active area of research during the last decades.

Although most contributions focus on the case where $\setDic$ contains a finite number of elements, several authors have recently placed the sparse-representation problem in the context of ``continuous'' dictionaries. In this framework, $\setDic$ is made up of an infinite (uncountable) number of elements continuously indexed by some parameter $\param$: 
\begin{align}\label{eq:def cont dico}
\setDic = \kset{\atom(\param)\in\kR^\dimobs}{\param\in \paramset}
\end{align}
where $\kfuncdef{\atom}{\paramset}{\kR^\dimobs}$ is a continuous function and $\paramset\subseteq\kR^\dimparam$. The generalization of the well-kwown Lasso problem to context of continuous dictionaries \eqref{eq:def cont dico} was considered in \cite{Bredies:2013lq}.
It involves the optimization with respect to a measure $\mesure$ belonging to the space of Radon measures over $\paramset$, say $\measureset(\paramset)$:
	\begin{equation}
		\label{eq:blasso}
		\mesure^\star\in \kargmin_{\mesure\in\calM(\Theta)}
		\tfrac{1}{2} \kvvbar{\vobs - \opMat\mesure}_2^2
		+ \lambda \kvvbar{\mesure}_{\mathrm{TV}},
	\end{equation}
	where $\opMat\mesure \triangleq \int \atom(\param) d\mesure(\param)$,
	and $\kvvbar{\cdot}_{\mathrm{TV}}$ is the so-called ``total variation''  norm, see \textit{e.g.}, \cite[Eq (7)]{Duval2015}.
We note that, despite of its apparent complexity (optimization over a space of measures), efficient procedures based on the conditional gradient method \cite[Section 2.2]{Bertsekas_book} have been proposed that operate on measures with finite support, see \textit{e.g.}, \cite{Boyd2017,Denoyelle2019aa}.

\paragraph{Low-rank approximation of matrices.}
Inspired by the success of the sparse model, several authors proposed to tackle inverse problems involving different types of low-dimensional structures. One striking example is the reconstruction of low-rank matrices considered by Candès \textit{et al} in \cite{Candes2009aa}.
More specifically, the authors addressed the problem of reconstructing a low-rank matrix $\mat\in\kR^{\dimmatA\times\dimmatB}$ from partial observations $\vobs\in\kR^{\dimobs}$ 
 as the solution of the following optimization problem:
	\begin{equation}
	\label{eq:low-rank-matrix}
		\mat^\star \in \kargmin_{\mat\in\kR^{\dimmatA\times \dimmatB}}
		\tfrac{1}{2} \kvvbar{\vobs - \opObs\mat}_2^2 + \lambda \kvvbar{\mat}_*.
	\end{equation}
Here $\kfuncdef{\opObs}{\kR^{\dimmatA\times\dimmatB}}{\kR^{\dimobs}}$ plays the role of some linear observation operator and $\kvvbar{\cdot}_*$ is the so-called ``nuclear'' norm defined as:
	\begin{equation}\label{eq:matrixnuclear}
		\kvvbar{ \mat }_* = \sum_{\ell=1}^{\min(\dimmatA,\dimmatB)} \sigma_\ell(\mat),
	\end{equation}
	where $\sigma_\ell(\mat)$ denotes the $\ell$th singular value of \(\mat\).
%
It can be shown that the rank of solutions to~\eqref{eq:low-rank-matrix} is related to \(\lambda\), \textit{i.e.} the nuclear norm induces low-rank solutions. The success of the method proposed by Candès \textit{et al.} revolves around: \textit{(i)} theoretical guarantees that the solution to~\eqref{eq:low-rank-matrix} may be the solution of the NP-hard low-rank approximation problem; \textit{(ii)} efficient solvers such as FISTA~\cite{Beck2009Fast} which rely on the closed-form expression of the proximal operator of the matrix nuclear norm.

\paragraph{Low-rank approximation of tensors.}
Given the success of convex-relaxation methods for the reconstruction of low-rank matrices, it is tempting to extend the ingredients of problem~\eqref{eq:low-rank-matrix} to the low-rank decomposition of $\dimtensor$-dimensional tensors.
To this end,
following~\cite{Lim2014Blind},  the tensor nuclear norm is defined as the gauge function of the convex hull of normalized rank-one tensors:
\begin{align}\nonumber
	\kvvbar{ \mat }_* = \inf \kset{\sum_{\idxsum<\infty} \kvbar{\cdec_\idxsum} }{\mat = \sum_{\idxsum< \infty} \cdec_\idxsum\, \vdecA_\idxsum \otimes \ldots \otimes \vdec^{(\dimtensor)}_\idxsum, \vdec^{(i)}_\idxsum\in\uS^{\dimsv_i}}.
\end{align}
where $\otimes$ denotes the tensor product and $\uS^{\dimsv_i}$ is the $\dimsv_i$-dimensional unit-sphere. Indexation \(\idxsum < \infty\) means the number of summands must be finite.
It is well-known that this definition is equivalent to~\eqref{eq:matrixnuclear} when \(t=2\)~\cite{Friedland2017}.

With this generalized definition (and a proper definition of $\opObs$), problem~\eqref{eq:low-rank-matrix} straightforwardly extends to the multi-dimensional tensorial case.
Unfortunately, as shown in~\cite{Friedland2017}, the evaluation of the nuclear norm is NP-difficult as soon as $\dimtensor>2$.
Although this seems to mean that (convex) nuclear norm regularized approximation~\eqref{eq:low-rank-matrix} is not easier to solve than (non-convex) tensor low-rank approximation, in this contribution we propose a Frank-Wolfe algorithm working on a related BLasso problem that tries to solve~\eqref{eq:low-rank-matrix} globally. We discuss research directions to hopefully provide appropriate assumptions for the global convergence of the algorithm, and show practical results for tensor rank selection.

\section{From low rank to sparsity}

The idea underlying the reformulation of the low-rank approximation problem as a sparse-representation problem is as follows:
any rank-$\nnz$ tensor can be rewritten as the sum of $\nnz$ rank-1 tensors; hence, rank-$\nnz$ tensors have $\nnz$-sparse representations in the (continuous) dictionary of rank-1 tensors.
In the framework of sparse representations in continuous dictionaries, one may define
the parameter set $\paramset$ as
	\begin{align}\label{eq:def:paramset}
		\paramset = \kset{(\vdecA,\ldots,\vdec^{(\dimtensor)})}{\vdec^{(i)}\in\uS^{\dimsv_i}}
	\end{align}
and the ``atom function'' as 
	\begin{align}\label{eq:def:atomfun}
		\kfuncdef{\atom}{\paramset\hspace*{1cm}}{\hspace*{1cm}\kR^{\dimobs}}[(\vdecA,\ldots,\vdec^{(\dimtensor)})][\opObs\kparen{\vdecA \otimes \ldots \otimes \vdec^{(\dimtensor)}}].
	\end{align}
With these definitions, the continuous dictionary $\setDic$ in \eqref{eq:def cont dico} corresponds to the image of the set of unit-norm rank-1 $\dimtensor$-dimensional tensors seen through operator $\opObs$.
A sparse decomposition of $\vobs$ in $\setDic$ can then be searched via the resolution of problem~\eqref{eq:blasso}.
The pertinence of this approach with respect to our initial regularized tensor-approximation problem~\eqref{eq:low-rank-matrix} is given in Proposition~\ref{prop:equivalence-solutions} below.

Before stating our result, we note that any solution $\mesure^\star$ of \eqref{eq:blasso} defines a $\dimtensor$-dimensional tensor as $\opMat\mesure^\star$, where $\opMat$ is defined from~\eqref{eq:def:paramset}-\eqref{eq:def:atomfun}.
Conversely, any solution $\mat^\star$ of \eqref{eq:low-rank-matrix} defines a discrete measure in $\measureset(\paramset)$ with mass $\sigma_\ell(\mat^\star)$ located at the corresponding singular vector, say $(\vdecA_\ell,\ldots,\vdec_\ell^{(\dimtensor)})$. Keeping this remark in mind, our result reads:
	\begin{proposition}\label{prop:equivalence-solutions}
	Consider problem~\eqref{eq:blasso} with $\opMat$ defined from~\eqref{eq:def:paramset}-\eqref{eq:def:atomfun}.
Any solution $\mat^\star$ of \eqref{eq:low-rank-matrix} defines a discrete solution of \eqref{eq:blasso}. Conversely, any discrete measure $\mesure^\star$ which is solution of \eqref{eq:blasso} defines a solution of \eqref{eq:low-rank-matrix}.
	\end{proposition}
\noindent
In other other words, \Cref{prop:equivalence-solutions} states that
all the solutions of~\eqref{eq:low-rank-matrix} can be found by identifying the discrete solutions of~\eqref{eq:blasso}.
Our proof revolves around the fact that the infimum in the nuclear norm is attained~\cite[Prop.~3.1]{Friedland2017}. In particular, it implies that there must exist discrete measures that are solutions to the BLasso problem.
We can therefore exploit the algorithmic strategies available for solving the BLasso problem to search for the solutions of~\eqref{eq:low-rank-matrix}.

\section{Algorithms}
A popular algorithm to solve the BLasso problem is the so-called `` Frank-Wolfe'' algorithm~\cite[Section 2.2]{Bertsekas_book}. We consider hereafter a generalized version of this procedure called ``Sliding Frank-Wolfe (SFW)'', see \cite[Chapter 4, Alg. 3]{Denoyelle2019aa}. It consists of three steps repeated until convergence. First, identify the best normalized atom by maximizing correlation with the residual.
Second, compute the optimal coefficients with fixed selected atoms. Finally, optimize both coefficients and atom parameters for a fixed sparsity level \( s \). In the tensor framework, this means alternately solving problems of the following form:
\begin{eqnarray}
&\displaystyle{\kargmax_{\vdec^{(i)}\in\uS^{\dimsv_i}}}\ \langle \opObs^{\ast}\mathbf{r}, \vdec^{(1)}\otimes \ldots \otimes \vdec^{(\dimtensor)} \rangle \label{eq:selection}	\\
&\displaystyle{\kargmin_{\mathbf{c}\in\kR^{s}}}\ \tfrac{1}{2}\|\vobs - \mathbf{D}\mathbf{c}\|_2^2 + \lambda \kvvbar{\mathbf{c}}_1	\label{eq:classo}\\
&\displaystyle{\kargmin_{\mathbf{c}\in\kR^s,\vdec^{(i)}_\ell\in\uS^{\dimsv_i}}} \tfrac{1}{2}\|\vobs - \opObs\sum_{\idxsum=1}^{s} \cdec_\idxsum \vdec^{(1)}_\ell\otimes \ldots \otimes \vdec^{(\dimtensor)}_\ell \|_2^2 + \lambda \kvvbar{\mathbf{c}}_1 \label{eq:sliding}
\end{eqnarray}
where \(\mathbf{r}\in\kR^{\dimobs}\) is the residual vector after a given number of iterations, matrix \(\mathbf{D}\) is computed straightforwardly from the set of identified atoms, and \(\opObs^{\ast}\) is the adjoint of \(\opObs\), often easy to compute in practice.

These three steps have very different levels of difficulty. Problem~\eqref{eq:classo} is a simple Lasso problem, for which efficient solvers already exist. Problem~\eqref{eq:sliding} is in principle as difficult as the original tensor factorization problem with low-rank constraint. Nevertheless, it can be shown that convergence of SFW is still ensured as soon as the output of step \eqref{eq:sliding} does not increase the BLasso cost function. Hence, problem~\eqref{eq:sliding} can be substituted by a few iterations of some local optimization procedures, see, \textit{e.g.}, \cite{Uschmajew2012Local}.

Finally, solving~\eqref{eq:selection} means finding the best rank-one
approximation of a tensor \( \opObs^{\ast}\mathbf{r} \), which is an NP-hard problem
in general~\cite{friedland2013best}. To the best of the author's knowledge, no
method is even proved to produce an estimate positively correlated with the
optimal solution in general~\cite{arous2019landscape}. On the other hand,
practical algorithms exist that empirically perform extremely
well~\cite{Friedland2017,da2015rank}. Moreover the Higher-Order Singular Value
Decomposition~\cite{Lathauwer2000Multilinear} provides an initialization which
guarantees quasi-optimality of the fitting error in Frobenius norm. We are currently working on
linking existing results on tensor best rank-one approximation with sufficient
conditions for the FW algorithm to converge globally. Practically, we
compute~\eqref{eq:selection} using the Alternating Least Squares algorithm, initialized randomly ten
times and picking the best result.

\section{Experiments}
Let us showcase the proposed SFW algorithm on a rank detection problem. Given a noised 3-dimensional data tensor \(\mathbf{T}=\mathbf{X} + \sigma\mathbf{N}\) with \(\mathbf{X}\) of rank \(\nnz=5\), we try to find the value of \(\nnz\) by solving the tensor version of~\eqref{eq:low-rank-matrix} for several values of \(\lambda \).
We set \( \opObs \) as the vectorization operator. The noise tensor has entries drawn from i.i.d. Gaussian distributions. Ground truth tensor \(\mathbf{X}\) is generated using factors drawn from i.i.d. Gaussian distributions with dimensions \(20\times21\times22\), and ground-truth \(\cdec_{\idxsum}\) are drawn from the absolute value of i.i.d. Gaussian distributions with mean \(0.5\). The noise level \( \sigma \) is set to \(0.005\).

\begin{figure}
	\centering
	\includegraphics[width=.8\columnwidth]{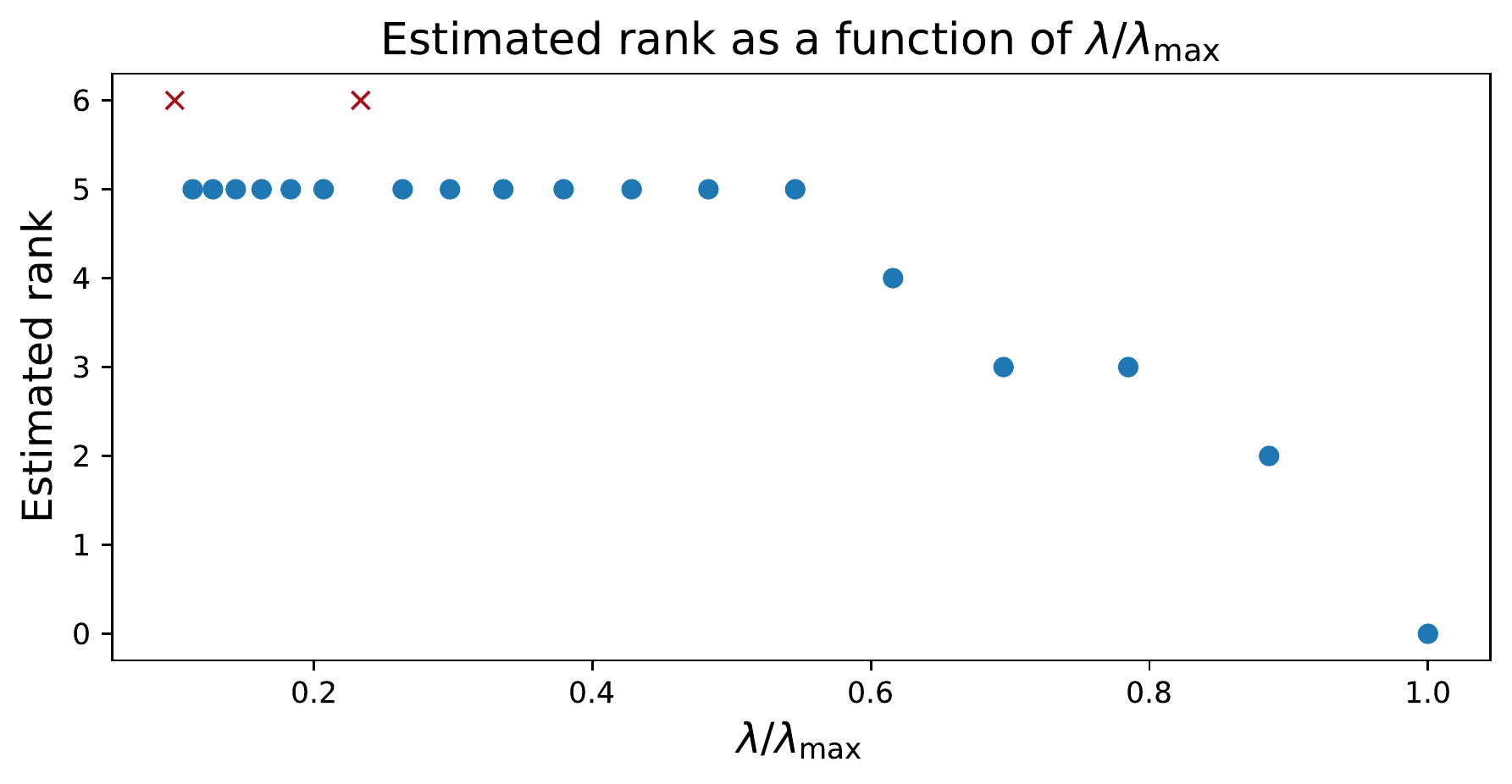}
	\caption{Tensor rank of the output of the proposed Sliding Frank-Wolfe algorithm, with \(\lambda\) varying from \(0\) to \(\lambda_{\max} \) as defined in~\cite{Denoyelle2019aa}. Red crosses are bad estimates, that might be improved by a better implementation of steps~\eqref{eq:selection} and~\eqref{eq:sliding} of the FW algorithm.
		\label{fig:estimate-rank}
	}
\end{figure}

Figure~\ref{fig:estimate-rank} shows the results for various values of \(\lambda\). We clearly observe proper rank detection, although there is a spurious error around \(\lambda = 0.25\lambda_{\max}\). Moreover, we observed that the atoms selected with various \(\lambda\) values are very closely related, which would confirm some kind of global convergence behavior, despite our implementation behind largely improvable. Note that within the proposed framework, we may also, among others, tackle missing data imputation or directly low-rank approximation.

\clearpage
\small
\bibliographystyle{elsarticle-num}

\end{document}

%% file: notations.tex
\usepackage{pgffor}
\foreach \x in {a,...,z}{
  \expandafter\xdef\csname bf\x \endcsname{\noexpand\ensuremath{\noexpand\mathbf{\x}}}
  \expandafter\xdef\csname bs\x \endcsname{\noexpand\ensuremath{\noexpand\boldsymbol{\x}}}
}

\foreach \x in {A,...,Z}{
  \expandafter\xdef\csname bs\x \endcsname{\noexpand\ensuremath{\noexpand\boldsymbol{\x}}}
  \expandafter\xdef\csname bf\x \endcsname{\noexpand\ensuremath{\noexpand\mathbf{\x}}}
  \expandafter\xdef\csname bb\x \endcsname{\noexpand\ensuremath{\noexpand\mathbb{\x}}}
  \expandafter\xdef\csname ds\x \endcsname{\noexpand\ensuremath{\noexpand\mathds{\x}}}
  \expandafter\xdef\csname cal\x \endcsname{\noexpand\ensuremath{\noexpand\mathcal{\x}}}
  \expandafter\xdef\csname fr\x \endcsname{\noexpand\ensuremath{\noexpand\mathfrak{\x}}}
}

\newtheorem{proposition}{Proposition}

\makeatletter

\usepackage{xcolor}
\definecolor{gris}{gray}{0.90}
\definecolor{gris25}{gray}{0.90}
\definecolor{americanrose}{rgb}{1.0, 0.01, 0.24}
\definecolor{bostonuniversityred}{rgb}{0.8, 0.0, 0.0}
\definecolor{shamrockgreen}{rgb}{0.0, 0.62, 0.38}
\definecolor{selectiveyellow}{rgb}{1.0, 0.73, 0.0}
\definecolor{royalblue}{rgb}{0.25, 0.41, 0.88}
\definecolor{ashgrey}{rgb}{0.7, 0.75, 0.71}
\definecolor{burgundy}{RGB}{159,29,53}
\definecolor{darkgreen}{RGB}{18,53,26}
\definecolor{lightblue}{RGB}{102,217,255}
\definecolor{fakeorange}{RGB}{255,140,102}
\definecolor{arylideyellow}{rgb}{0.91, 0.84, 0.42}
\definecolor{bananayellow}{rgb}{1.0, 0.88, 0.21}
\definecolor{gris_f}{gray}{0.35}
\definecolor{bordure}{rgb}{0.09,0.17,0.68}
\definecolor{aquamarine}{rgb}{0.5, 1.0, 0.83}
\definecolor{apricot}{rgb}{0.98, 0.81, 0.69}
\definecolor{babyblue}{rgb}{0.54, 0.81, 0.94}
\definecolor{uipoppy}{RGB}{225, 64, 5}
\definecolor{uipaleblue}{RGB}{96,123,139}
\definecolor{uiblack}{RGB}{0, 0, 0}
\definecolor{decoda}{RGB}{0,153, 0}
\definecolor{lightgreen}{rgb}{0.56, 0.93, 0.56}
\definecolor{blue_f}{rgb}{0.2, 0.2, 0.6}
\definecolor{cinnamon}{rgb}{0.82, 0.41, 0.12}
\definecolor{darkpastelgreen}{rgb}{0.2, 0.75, 0.24}
\definecolor{drab}{rgb}{0.59, 0.44, 0.09}

\newcommand{\dimobs}{m}
\newcommand{\nnz}{k}
\newcommand{\dimsv}{n}
\newcommand{\atom}{\mathbf{d}}
\newcommand{\param}{\boldsymbol{\theta}}
\newcommand{\paramset}{\Theta}
\newcommand{\dimparam}{d}
\newcommand{\measureset}{\mathcal{M}}
\newcommand{\dimmatA}{n_1}
\newcommand{\dimmatB}{n_2}

\newcommand{\idxsum}{\ell}

\newcommand{\cdec}{c}
\newcommand{\vdec}{\mathbf{u}}
\newcommand{\vdecA}{\vdec^{(1)}}

\newcommand{\uS}{\mathbb{S}}    
\newcommand{\dimtensor}{t}
